\documentclass[aip,apl,reprint]{revtex4-1}
\usepackage{multirow}
\usepackage{array}
\usepackage[colorinlistoftodos]{todonotes}
\usepackage{amsmath}
\usepackage{amssymb}
\usepackage{bm}
\usepackage{color}
\usepackage{graphicx}
\usepackage[latin1]{inputenc}
\usepackage[english]{babel}
\usepackage{color, soul}
\usepackage{xcolor}

\usepackage{ulem}
\begin{document}

%\preprint{AIP/123-QED}

\title{Direct measurement of interfacial Dzyaloshinskii-Moriya interaction at the % \sout{few-layer}
MoS$_{\rm 2}$/Ni$_{80}$Fe$_{20}$ interface}

\author{Akash Kumar,\textit{$^{1}$}$^{\dag}$, Avinash Kumar Chaurasiya,\textit{$^{2}$}$^{\dag}$ Niru Chowdhury,\textit{$^{1}$} Amrit Kumar Mondal,\textit{$^{2}$} Rajni Bansal,\textit{$^{1}$} Arun Barvat,\textit{$^{3}$} 
Suraj P Khanna,\textit{$^{3}$}
Prabir Pal,\textit{$^{4}$} Sujeet Chaudhary,\textit{$^{1}$} Anjan Barman,\textit{$^{2}$} and P.~K.~Muduli}
\email{muduli@physics.iitd.ac.in\\
$\dag$ These authors contributed equally to this work}

\affiliation{$^{1}$Thin Film Laboratory, Department of Physics, Indian Institute of Technology Delhi, Hauz Khas, New Delhi-110016, India\\$^{2}$Department of Condensed Matter Physics and Material Sciences, S. N. Bose National Centre for Basic Sciences, Block JD, Sec. III,
Salt Lake, Kolkata 700106, India\\ $^{3}$CSIR-National Physical Laboratory, Dr.~K.~S.~Krishnan Road,~New Delhi-110012,~India\\ 
%$^{4}$Academy of Scientific and Innovative Research (AcSIR), CSIR-HRDC Campus, Ghaziabad 201002, India\\
$^{4}$CSIR-Central Glass and Ceramic Research Institute, 196, Raja S. C. Mullick Road, Kolkata 700032, India
}

\begin{abstract}
 
We report on a direct measurement of sizable interfacial Dzyaloshinskii-Moriya interaction (iDMI) at the interface of two-dimensional transition metal dichalcogenide (2D-TMD), MoS$_{\rm 2}$ and Ni$_{80}$Fe$_{20}$ (Py) using Brillouin light scattering spectroscopy. A clear asymmetry in spin-wave dispersion is measured in MoS$_{\rm 2}$/Py/Ta, while no such asymmetry is detected in the reference Py/Ta system. A linear scaling of the DMI constant with the inverse of Py thickness indicates the interfacial origin of the observed DMI. We further observe an enhancement of DMI constant in three to four layer MoS$_{\rm 2}$/Py system (by 56$\%$) as compared to 2 layer MoS$_{\rm 2}$/Py which is caused by a higher density of MoO$_{\rm 3}$ defect species in the case of three to four layer MoS$_{\rm 2}$. The results open possibilities of spin-orbitronic applications utilizing the 2D-TMD based heterostructures.

\end{abstract}

\maketitle

%\section{Introduction}
Dzyaloshinskii-Moriya interaction (DMI)~\cite{Dzyloshinskii1957, Moriya1960} favors perpendicular alignment of neighboring spins in a ferromagnetic material. In competition with Heisenberg exchange interaction, the DMI can lead to the formation of stable chiral spin textures such as N\'eel type domain walls or skyrmions,~\cite{roessler2006Skyrmtheory,bode2007chiral} which are potential candidates for memory and logic applications due to their efficient current-driven motion and smaller size.~\cite{parkin2008magnetic,parkin2015,Tomasello2014} These chiral structures were first observed in bulk non-centrosymmetric B20 magnetic materials such as MnSi,~\cite{muhlbauer2009skyrmion} and FeGe~\cite{yu2011nearRTskyrm} due to the inherent broken inversion symmetry in these materials. 
In ferromagnet/heavy metal (FM/HM) heterostructures, the large spin-orbit coupling of the HM and the broken inversion symmetry at the interface result in an interfacial Dzyaloshinskii-Moriya interaction (iDMI).~\cite{fert1980prl,Yang2015DMI} More recently, stable skyrmions are observed in thin-film heterostructures of FM/HM bilayers due to the presence of an iDMI.~\cite{husain2019observation,jiang2017skyrmions} The FM/HM bilayer structures are technologically advantageous since they provide the opportunity to control and manipulate skyrmion/domain walls using current-induced spin-orbit torques present in these systems. Hence, the quantification of DMI constant is important both for fundamental physics as well as for designing efficient FM/HM systems for applications. Several direct and indirect measurement techniques such as Brillouin light scattering (BLS) spectroscopy,~\cite{nembach2015DMI,di2015apl,Di2015PRL,Belmeguenai2015SDMI,Stashekvich2015DMI} domain wall velocity,~\cite{torrejon2014DWmotion} magnetic force microscope measurements,~\cite{bacani2019measure} asymmetric hysteresis loop method,~\cite{han2016perspectives} etc.
have been developed to quantify the strength of DMI. However, the wave vector (\textit{k}) dependent BLS measurements have been established as a direct and very reliable method of measuring the strength of DMI.~\cite{nembach2015DMI,Di2015PRL,chaurasiya2016sr,Chaurasiys2018PRApp}

A large section of BLS measurement of DMI are focused to the FM/HM systems, namely, Py/Pt,~\cite{nembach2015DMI} Co/Pt,~\cite{Belmeguenai2015SDMI} CoFeB/W,~\cite{chaurasiya2016sr} CoFeB/Ta,\cite{Chaurasiys2018PRApp} CoFe/Pt,~\cite{Belmeguenai2016DMI} 
and Ta/Co$_{2}$FeAl/MgO~\cite{husain2019observation} systems. More recently higher DMI strength is reported for superlattices of [Co/Pd(111)]~\cite{Davydenko2019PdDMI} and [Ir/Fe/Co/Pt].~\cite{raju2019evolution} Apart from conventional HMs, recent results have shown the presence of iDMI~\cite{Akash2019Gr,yang2018GrDMI,ajejas2018unraveling} in Graphene/FM bilayer structures. Generally, DMI scales with spin-orbit coupling strength in the material in contact with the FM layer.~\cite{fert1980prl,Yang2015DMI} Hence, the presence of DMI in the Graphene-based heterostructure was surprising, given the low intrinsic spin-orbit coupling strength of Graphene. These results were explained based on the Rashba effect originated at the Graphene-FM interface~\cite{yang2018GrDMI,ajejas2018unraveling} or due to the extrinsic spin-orbit coupling at the Graphene-FM interface.~\cite{Akash2019Gr} 
However, the magnitude of the DMI parameter for these Graphene-based two-dimensional (2D) material-heterostructures is significantly lower compared to the FM/HM systems. In order to enhance the DMI parameter in the 2D material-heterostructures, 2D transition metal dichalcogenides (TMDs) materials are very promising, since they possess larger intrinsic spin-orbit coupling compared to Graphene. In addition, the lack of inversion center~\cite{xiao2012coupled} in the crystal structure of TMDs provides immense advantage, which is already utilized to obtain unconventional spin-orbit torques in 2D TMDs/FM systems.~\cite{Macneil2017NP_WTe2,shao2016nanoletters,guimaraes2018spin}

In this work, we report on the direct observation of a sizable DMI in a reasonably large area and few layer-MoS$_{\rm 2}$/Ni$_{80}$Fe$_{20}$ (Py) heterostructures using BLS measurements. MoS$_{\rm 2}$ is a layered TMD having large intrinsic spin-orbit coupling which leads to a giant spin splitting due to the absence of inversion symmetry.~\cite{Zhu2011DFT_TMD} The measured DMI in this system is found to be larger compared to Graphene/Py system~\cite{Akash2019Gr,yang2018GrDMI,ajejas2018unraveling} and having comparable magnitude as that of the FM/HM bilayer systems for the similar thickness of FM layer.~\cite{nembach2015DMI, chaurasiya2016sr,Chaurasiys2018PRApp,Belmeguenai2016DMI} The linear dependence of the DMI with the inverse of Py thickness and the absence of DMI in the reference sample indicate that the DMI in this system originates from the interface of MoS$_{\rm 2}$ and Py layer. We found larger DMI for three to four layer (L) MoS$_{\rm 2}$/Py system compared to the 2L MoS$_{\rm 2}$/Py, which is attributed to a higher density of MoO$_{\rm 3}$ defect species in the 3-4L MoS$_{\rm 2}$.

%\section{Experimental}

We use large-area 2L and 3-4L MoS$_{\rm 2}$ grown on Si/SiO$_{\rm 2}$ substrates by pulsed laser deposition (PLD) technique. The substrates are cleaned with standard chemical cleaning procedures followed by annealing at 700~$^\circ$C in a vacuum chamber %with a base pressure of $2\times10^{-6}$ Torr, 
for out-gassing of all impurities before deposition of MoS$_{\rm 2}$ thin-films. The MoS$_{\rm 2}$ thin films are then deposited using a 248 nm KrF excimer laser source at 5 Hz repetition frequency, 20 ns pulse width and 0.5~$\mu$J/m$^2$ energy density. The base pressure of $6\times10^{-6}$~Torr is maintained throughout the deposition process. The thickness of MoS$_{\rm 2}$ was varied by the number of laser shots. The laser shots of 40 and 60 are optimized for 2L (layer) and 3-4L MoS$_{\rm 2}$ growth, respectively. More details of MoS$_{\rm 2}$ growth along with the characterization of bare MoS$_{\rm 2}$ can be found in Ref~\onlinecite{Barvat2017}. 
The Py and the Ta layers were grown in a separate DC magnetron sputtering chamber with a base vacuum of $2\times10^{-6}$ Torr after cleaning MoS$_{\rm 2}$ samples with acetone and isopropyl alcohol. The Py thickness ($t_{\rm Py}$) was varied from $3-20$~nm while the Ta capping layer thickness was fixed at 2~nm. The bilayer metallic films are deposited at room temperature with a working pressure of $1\times10^{-3}$ Torr. A set of reference samples \textit{i.e.,} Py ($t_{\rm Py}$)/Ta (2 nm) were also prepared simultaneously for comparison. The growth rate of Py and Ta thin-films were maintained at 1.43 $\rm \AA$/s and 1.83 $\rm \AA$/s, respectively. %The Ta layer acts as a protective layer, which is primarily self-oxidized and does not provide any iDMI at the Py/Ta interface. 

The BLS measurements have been carried out in Damon-Eshbach (DE) geometry using a Sandercock type 3+3 pass Tandem Fabry-P\'erot interferometer. Conventional 180{\textdegree} back scattered protocol along with the tool for wave vector selectivity was followed. As BLS relies on the inelastic light scattering process where total momentum is conserved in the plane of the thin film, Stokes (anti-Stokes) peaks are observed in the BLS spectra which correspond to the creation (annihilation) of magnons with momentum $k$=$(4\pi/\lambda)$sin$\theta$, 
where $\lambda$= 532 nm in our case and $\theta$ is the angle of incidence of the laser beam. A well defined BLS spectrum was obtained after counting photons for several hours.~\cite{chaurasiya2016sr,Akash2019Gr} In the DE geometry, the external magnetic field is applied perpendicular to the plane of incidence of the laser beam. This geometry allows the probing of the spin-waves propagating along the in-plane direction perpendicular to the externally applied field. More details of the BLS measurements can be found elsewhere.~\cite{Akash2019Gr, chaurasiya2016sr,Chaurasiys2018PRApp}

 \begin{figure} [t!]
\centering
\includegraphics[width=\columnwidth]{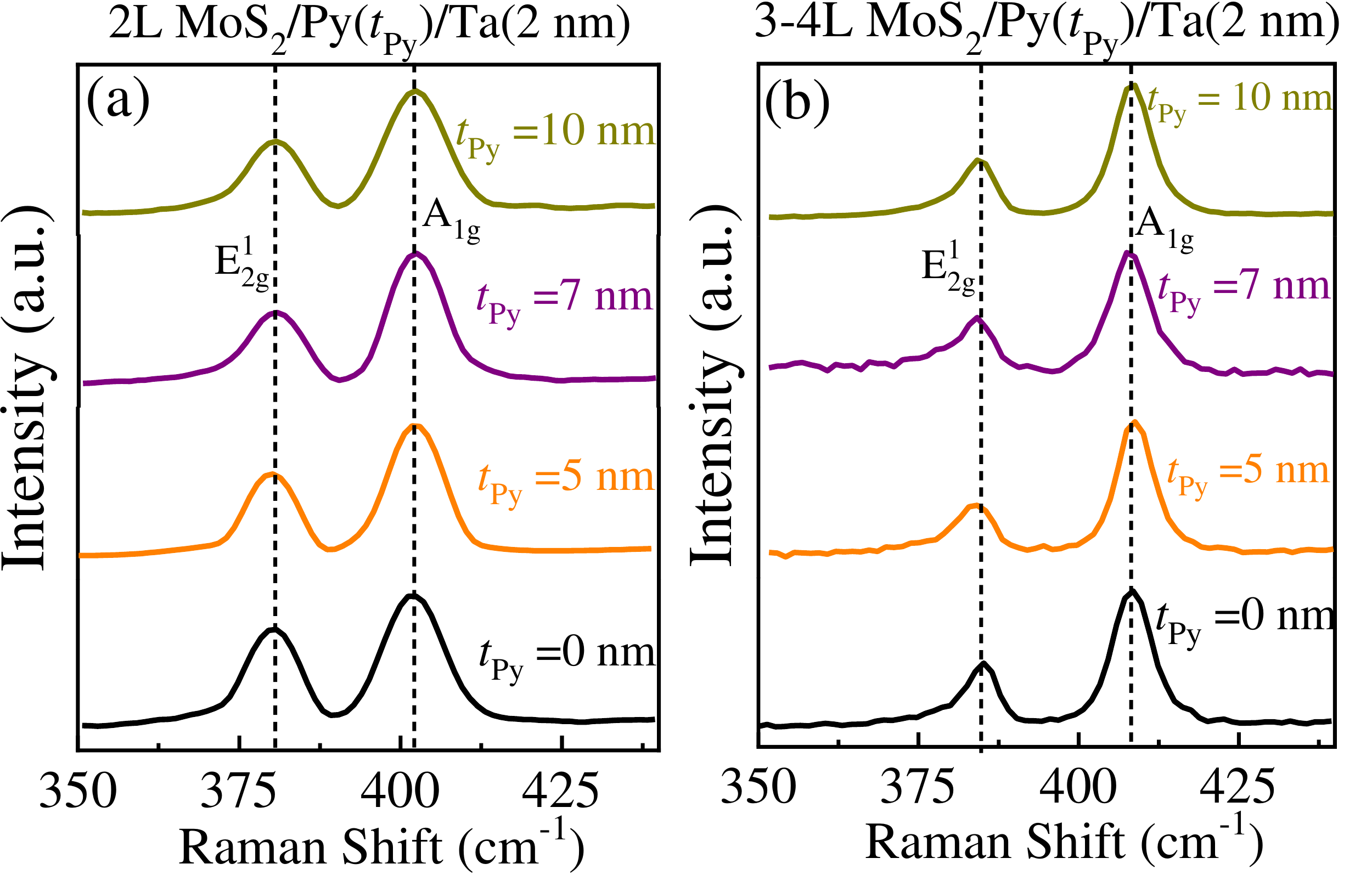}
\caption{\label{fig:RAMAN} Raman measurements of (a) 2L MoS$_{\rm 2}$/Py and (b) 3-4L MoS$_{\rm 2}$/Py with varying Py thickness. The dashed lines represent the peak positions of E$_{2g}^{1}$ and A$_{1g}$ peaks.
}
\end{figure}
 
%\section{Results and Discussion}
Figure~\ref{fig:RAMAN} shows the Raman measurements performed on both bare 2L and 3-4L MoS$_{\rm 2}$ and after deposition of Py/Ta for varying thicknesses of Py. Two peaks observed at $\sim$383 cm$^{-1}$ and $\sim$408 cm$^{-1}$ are the fingerprints of MoS$_{\rm 2}$ and correspond to the E$_{\rm 2g}^{1}$ and $A_{\rm 1g}$ modes, respectively. The observed difference in peak position of E$_{\rm2g}^{1}$ and $A_{\rm 1g}$ modes, denoted as $\delta$ is widely used as a reliable method of determining layer thickness for MoS$_{\rm 2}$.~\cite{li2011adfm} With the increasing number of layers, the E$_{\rm2g}^{1}$ mode shifts to lower frequencies, while the $A_{\rm 1g}$ mode shifts to higher frequencies. For the Raman spectra shown in Fig.~\ref{fig:RAMAN}(a), $\delta=(22.0\pm 0.3)$~cm$^{-1}$ is consistent with 2 layers of MoS$_{\rm 2}$, while for the Raman spectra shown in Fig.~\ref{fig:RAMAN}(b), we found $\delta=(24.6\pm 0.9)$~cm$^{-1}$ which corresponds to three to four layers of MoS$_{\rm 2}$. 
The difference $\delta$, we report here, is the average of all the 2L and 3-4L MoS$_{\rm 2}$ samples, since no significant differences in the values of $\delta$ were observed as a function of the Py thickness. These Raman measurements also show that the quality of MoS$_{\rm 2}$ layer remains unchanged even after the deposition of Py and Ta capping layers, as we do not see any systematic change in the linewidth of the Raman peaks with Py thickness. We also do not observe any additional peaks after deposition of Py and Ta, which is often observed when the disorder is introduced into MoS$_{\rm 2}$.~\cite{mignuzzi2015prb} More details of Raman mapping measurements~\cite{bansal2019MoS2} confirm that the MoS$_{\rm 2}$ layers are large area and the difference $\delta$ is maintained throughout the sample size of 2$\times$5~mm$^2$.

 \begin{figure*} [t!]
\centering
\includegraphics[width=1.8\columnwidth]{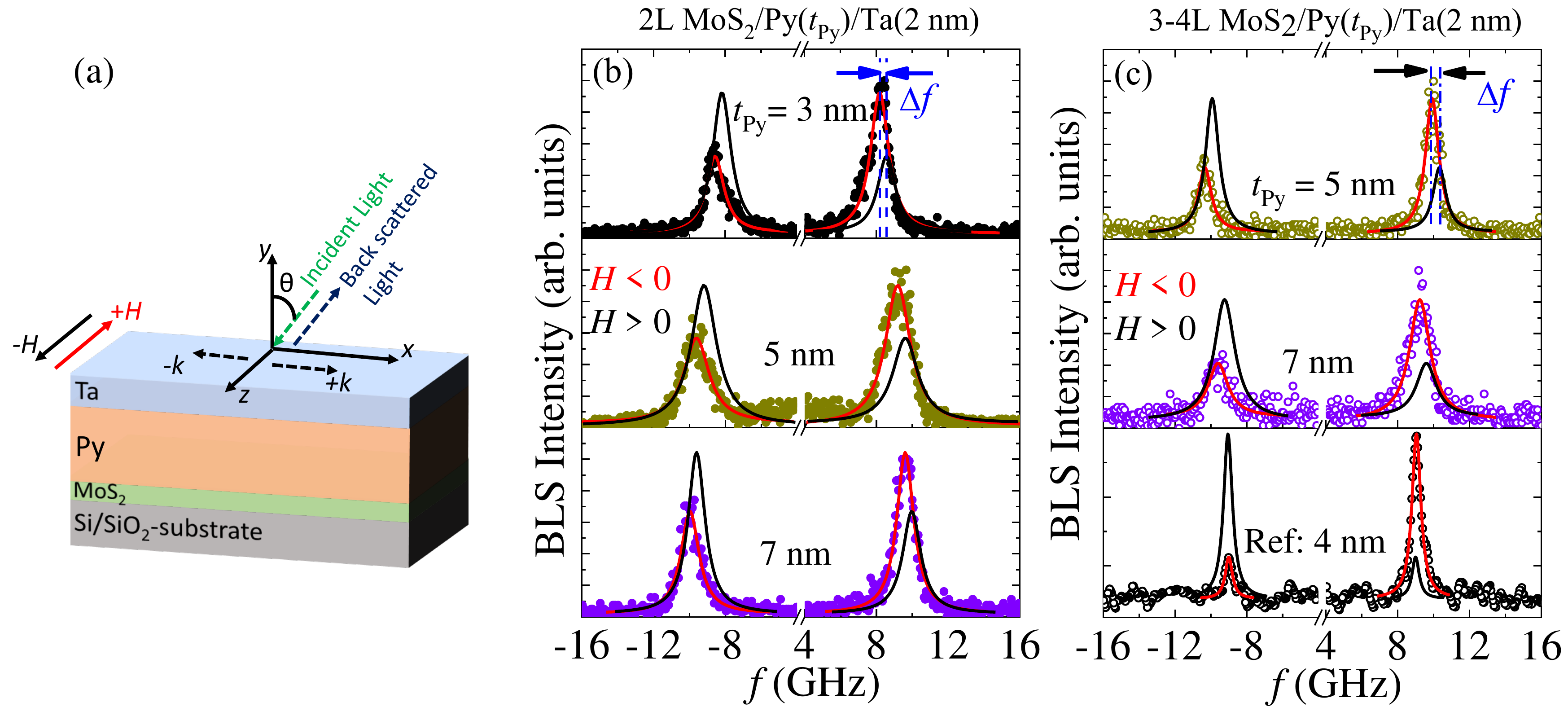}
\caption{\label{fig:BLS1} (a) Schematic of thin-film stacks with BLS measurement geometry. Measured BLS spectra from (b) 2L MoS$_{\rm 2}$/Py/Ta and (c) 3-4L MoS$_{\rm 2}$/Py/Ta samples, respectively, with varying Py thicknesses [lowest panel of (c) shows the BLS spectrum from reference sample; Py (4 nm)/Ta (2 nm)]. The symbols represent the measured BLS spectra while the red lines show fits with Lorentzian function. (black lines are counter-propagating fits). }
\end{figure*}

Figure~\ref{fig:BLS1}(a) shows the schematic of the thin film stack and the geometry used for obtaining BLS spectra in our samples. Here, the magnetic field is applied in the \textit{z}-direction and \textit{k}-vector is changed by varying the incident angle ($\theta$). Examples of measured BLS spectra for 2L and 3-4L MoS$_{\rm 2}$/Py ($t_{\rm Py}$)/Ta (2 nm) samples are shown in Fig.~\ref{fig:BLS1}(b) and (c), respectively. Here, the measurements were performed for %the wave vector of the spin-waves;
$k = 15.8$~rad/$\mu$m with in-plane magnetic field of $H = 1$~kOe. The lower panel of Fig.~\ref{fig:BLS1}(c) shows the BLS spectrum obtained for a reference sample without the MoS$_{\rm 2}$ layer \textit{i.e.,} Py (4 nm)/Ta (2 nm) sample (for $k = 11.4$~rad/$\mu$m and $H = 1$~kOe). The BLS spectra are well fitted with the Lorentzian function (fit shown with solid lines) to get the spin-wave frequency value (\textit{f}). An asymmetry in the frequencies of Stokes and anti-Stokes peaks is present in all samples where both 2L and 3-4L MoS$_{\rm 2}$ is interfaced with Py while it is absent in the case of the reference sample without MoS$_{\rm 2}$. The frequency difference ($\Delta f$) between Stokes and anti-Stokes peak position, which quantifies the strength of DMI, in counter-propagating spin-waves, is found to decrease with the increase in the thickness of Py layer. This observation confirms the interfacial origin of DMI in our system.~\cite{Belmeguenai2015SDMI,nembach2015DMI} Furthermore, $\Delta f$ is found to be negative for the positive applied magnetic field, revealing a negative sign of DMI in these samples.~\cite{Di2015PRL,Belmeguenai2015SDMI} This sign is consistent with Pt-based system with similar HM/FM stack ordering.~\cite{nembach2015DMI,Belmeguenai2016DMI} However, the sign of DMI in MoS$_{\rm 2}$/Py based heterostructures is opposite as compared to that of Graphene/Py, where the same stack order was used.~\cite{Akash2019Gr} While the reason behind this opposite sign is not understood completely, the higher electronegativity of sulfur atoms in MoS$_{\rm 2}$ may be responsible. Similar results were observed in the HM/FM system for which the electronegativity of HM is known to play an important role.~\cite{torrejon2014DWmotion}

%The change in the size and sign of the DMI with the heavy-metal elements may be related to the change in the charge localization of the interface atoms38, which has been reported to change the sign of the Rashba spin splitting at metal alloy surfaces39.

\begin{figure}[b!]
\centering
\includegraphics[width=\columnwidth]{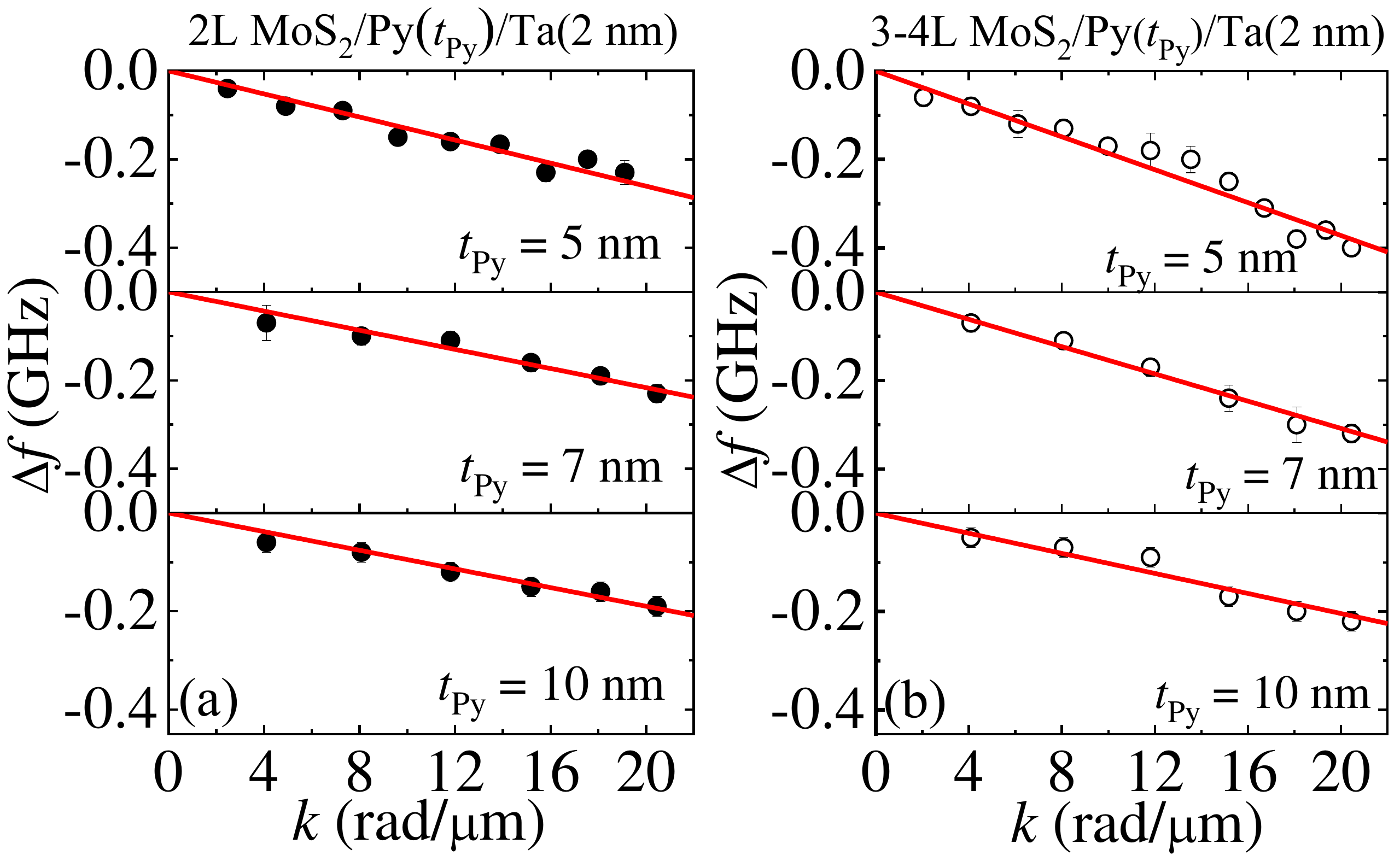}
\caption{\label{fig:BLS2} Average frequency asymmetry ($\Delta f$) between Stokes and anti-Stokes lines in BLS measurements as a function of wave vector \textit{k} of the spin-wave for (a) 2L MoS$_{\rm 2}$/Py and (b) 3-4 L MoS$_{\rm 2}$/Py samples at different Py thicknesses.}
\end{figure}

The $k$-dependent BLS measurement has been performed by changing the angle of incidence of the laser beam to the sample. Average frequency asymmetry ($\Delta f$) can be determined using:
\begin{equation}\label{delF}
\Delta f= \frac{[f(-k,M_{\rm z})-f(k,M_{\rm z})]-[f(-k,-M_{\rm z})-f(k,-M_{\rm z})]}{2} {,}
\end{equation}
where, $M_{\rm z}$ is magnetization in the direction of applied magnetic field. 
The strength of DMI constant ($D$) can be calculated using:~\cite{di2015apl}
\begin{equation}\label{DMI}
\Delta f = \frac{2\gamma}{\pi M_{\rm S}}Dk %+ \Delta \epsilon
\end{equation}
where, $\gamma$ (=1.85$\times10^{2}$~GHz/T) is the gyromagnetic ratio for Py thin-films. $M_{\rm S}$ is the saturation magnetization, which was determined from thickness dependent ferromagnetic resonance (FMR) data and magnetometry measurements.~\cite{bansal2019MoS2}
Figure~\ref{fig:BLS2} (a) and (b) shows frequency difference measured at different $k$ values for samples with various Py thickness for both 2L and 3-4L MoS$_{\rm 2}$, respectively. The maximum value of $\Delta f $ is found to be around 0.3 GHz for 2L-MoS$_{\rm 2}$/Py (5 nm) sample and 0.4~GHz for 3-4L MoS$_{\rm 2}$/Py (5 nm) sample. % $\Delta \epsilon\approx$0.04 GHz measured in our system is very small compared to the DMI (first) term.
The $\Delta f$ is also found to increase with the decrease of FM thickness indicating the dominance of DMI induced frequency non-reciprocity as opposed to the interfacial out-of-plane anisotropy induced frequency non-reciprocity which should vary linearly~\cite{Stashekvich2015DMI} or quadratically~\cite{gladiiprb2016} with the thickness of the FM layer.

\begin{figure}[t!]
\centering
\includegraphics[width=7.0cm]{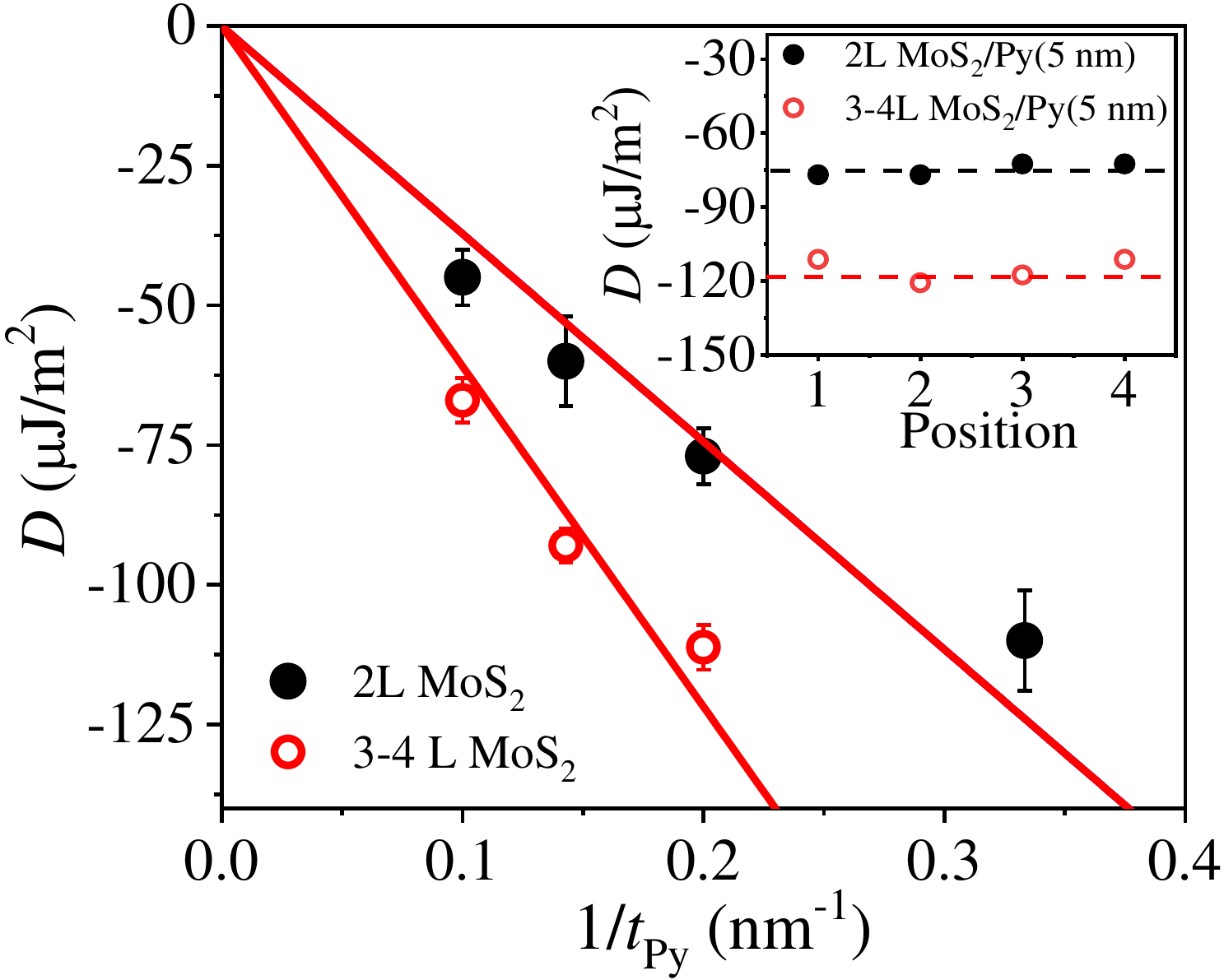}
\caption{\label{fig:FMthickness} Extracted DMI constant (\textit{D}) vs. inverse of Py thickness (1/$t_{\rm Py}$) for 2L and 3-4L MoS$_{\rm 2}$/Py samples.
Inset shows spatial dependence of \textit{D} for 2L and 3-4L MoS$_{\rm 2}$/Py (5 nm) samples. Symbols represent the observed data and the dashed lines represent the average values.}
\end{figure}

As the right hand side in Eq.~(\ref{DMI}) is linear with both $D$ and $k$, the slope of the linear fit to $\Delta f $ vs. $k$ can be used to determine the DMI constant accurately. The data (black circles) in Fig.~\ref{fig:BLS2} was fitted (solid red line) using Eq.~(\ref{DMI}) to obtain the effective DMI constant. 
We can expect a linear behaviour of \textit{D} with the inverse of FM layer thickness for the purely interfacial origin of DMI.~\cite{nembach2015DMI,Belmeguenai2015SDMI,chaurasiya2016sr} Figure~\ref{fig:FMthickness} shows the measured value of $D$ as a function of inverse of Py thickness (1/$t_{\rm Py}$) for both 2L (black solid circles) and 3-4L MoS$_{\rm 2}$/Py (red open circles) samples. The solid lines in Fig.~\ref{fig:FMthickness} shows linear fit to the data using the following equation:~\cite{Belmeguenai2015SDMI}
\begin{equation}\label{eq:D}
D= \frac{D_{\rm S}}{t_{\rm Py}}
\end{equation}

Here, $D_{\rm S}$ is the strength of the surface DMI parameter, which is independent of FM thickness. It is found that the $D_{\rm S} = (-0.39 \pm 0.03)$~pJ/m for 2L MoS$_{\rm 2}$/Py and $(-0.61 \pm 0.04)$~pJ/m for 3-4L MoS$_{\rm 2}$/Py. The observed values are larger than the previously reported results on monolayer Graphene ($D_{\rm S}$=0.19 pJ/m)~\cite{Akash2019Gr} and are comparable %(differing only by factor 2) 
to widely studied Pt heavy metal-based system ($D_{\rm S}=-1.7$ pJ/m).~\cite{Belmeguenai2015SDMI,nembach2015DMI} This sizable DMI value shows that MoS$_{\rm 2}$ can be a very important material for spintronic and magnonic applications.

We also measure spatial variation of DMI in our 2L and 3-4L MoS$_{\rm 2}$/Py bilayers, since we use large area MoS$_{\rm 2}$. Inset of Fig.~\ref{fig:FMthickness}~shows the spatial dependence of \textit{D} observed at different positions of the sample. The sample size was approximately 2$\times$5~mm$^2$ and the measurements are performed at the center and three other corners of the sample with approximate separation of $2-2.5$~mm. The laser spot width was about 50 $\mu$m; hence these measurement positions are not overlapped with each other but significantly apart.  The value of \textit{D} for both 2L and 3-4L MoS$_{\rm 2}$/Py does not show significant spatial variation with respect to the various positions of the samples as revealed by BLS measurement. The maximum variation is around 8.5 \% for 3-4L MoS$_{\rm 2}$/Py and 6.1 \% for 2L MoS$_{\rm 2}$/Py. Hence, higher DMI observed for in 3-4L MoS$_{\rm 2}$/Py compared to 2L MoS$_{\rm 2}$/Py is valid in the entire large area of the sample.

We found a larger $D_{\rm S}$ for both 2L and 3-4L MoS$_{\rm 2}$/Py samples compared to Graphene/Py,~\cite{Akash2019Gr} which can be attributed to larger spin-orbit coupling (SOC) of MoS$_{\rm 2}$. 
Furthermore, the value of $D_{\rm S}$ for 3-4L MoS$_{\rm 2}$/Py sample is found significantly higher (56$\%$) than the 2L MoS$_{\rm 2}$/Py. However, this result is counter-intuitive due to the following reason. The strength of DMI in FM/non-magnet system scales with SOC strength of the non-magnetic material in contact with the FM. It is known that intrinsic SOC in MoS$_{\rm 2}$, which acts as a non-magnetic layer in our study, decreases with an increase in the number of layers.~\cite{chang2014MoS2thickness} Hence, we expect a larger SOC in 2L MoS$_{\rm 2}$ than 3-4L MoS$_{\rm 2}$, as observed in our photo-luminescence measurements (not shown), which showed smaller spin splitting for 3-4L MoS$_{\rm 2}$.~\cite{bansal2019MoS2} Hence, intrinsic SOC cannot explain the observed results. However, our earlier XPS data~\cite{Barvat2017} and Raman mapping~\cite{bansal2019MoS2} on these MoS$_{\rm 2}$ samples show the presence of higher sulfur vacancies in 3-4L MoS$_{\rm 2}$ layer. In the XPS data, it is also observed that the presence of sulfur vacancies was primarily due to the formation of MoO$_3$ species in the case of 3-4L MoS$_{\rm 2}$. The large number of MoO$_{3}$ species in the case of 3-4L MoS$_{\rm 2}$ may be caused by the decomposition of MoS$_2$ target for a higher number of laser shots used during PLD growth, which will increase the number of sulfur vacancies and promote the formation of MoO$_{3}$ species. As previously shown that larger electronegativity leads to larger DMI.~\cite{torrejon2014DWmotion} Since oxygen has a higher electronegativity of 3.44, compared to sulfur for which the electronegativity is 2.58. Hence, we argue that higher value of \textit{D} observed in 3-4L MoS$_{\rm 2}$/Py interface is due to the formation of local defects of MoO$_3$. In BLS measurement, we measure volume averaged DMI constant, which can enhance due to the higher number of 
MoO$_{\rm 3}$ defect species in the case of 3-4L MoS$_{\rm 2}$/Py. 
%Furthermore, these results are consistent with our earlier spin pumping measurements.~\cite{bansal2019MoS2} An enhancement in spin mixing conductance was observed in 3-4L MoS$_{\rm 2}$/Py as compared to 2L MoS$_{\rm 2}$/Py interface. Since the spin mixing conductance also depends on SOC,~\cite{Chen2015SP} we believe that defect-induced extrinsic SOC at the interface is the primary origin of iDMI in our system while an intrinsic SOC-based iDMI can also not be ruled out. However, its value should not vary from 2L to 3-4L MoS$_{\rm 2}$. It is also non-trivial to isolate the intrinsic and extrinsic contributions in our present experiment.

%\section{Conclusion}
In conclusion, we have shown a direct measurement of sizable DMI constant for 2L and 3-4L MoS$_{\rm 2}$/Py system using BLS measurements. The FM thickness dependence of DMI reveals the dominating interfacial origin of DMI in the studied system. We have observed a larger interfacial DMI in 3-4L MoS$_{\rm 2}$/Py system as compared to 2L MoS$_{\rm 2}$/Py, which can be correlated with the higher density of MoO$_{\rm 3}$ defect species at the interface for 3-4L MoS$_{\rm 2}$/Py. 
The maximum value of DMI constant obtained is comparable to that of previously reported values for Pt-based FM/HM heterostructures for the similar thickness of the FM layer. Hence, these results show the possibility of stabilizing chiral spin textures and their manipulation using already emerging unconventional spin-orbit torques in the 2D materials/FM system. Further, the iDMI at MoS$_{\rm 2}$/Py interface can be controlled by engineering defects, which opens another pathway for the development of spintronic devices using 2D-TMDs. 

The data that supports the findings of this study are available within the article.

\begin{acknowledgments}
The partial support from the Ministry of Human Resource Development under the IMPRINT program (Grant no: 7519 and 7058), the Department of Electronics and Information Technology (DeitY), and Science and Engineering Research Board (SERB) under the Core Research Grant (CRG)
(grant no: SERB/F/12383/2018-2019) are gratefully acknowledged. A.~B. acknowledges S. N. Bose National Centre for Basic Sciences for financial assistance under grant no. SNB/AB/18-19/211. A.~K. acknowledges support from the Council of Scientific and Industrial Research (CSIR), India, while A.K.C. acknowledges DST, Government of India for INSPIRE fellowship (Grant No.
IF150922).
\end{acknowledgments}

\bibliography{References.bib}

\end{document}